\documentclass[aps,prc,twocolumn,showpacs]{revtex4}
\usepackage{amssymb}
\usepackage{amsmath}
\usepackage{graphicx}
\usepackage{lscape}
\usepackage{booktabs}
\usepackage{epsfig}

\allowdisplaybreaks

\begin{document}

\title{Reaction $e^+e^-\to\bar DD$ and $\psi'$ mesons}
\author{A. Limphirat$^{1,2}$, W. Sreethawong$^{1,2}$,
%\footnote{Corresponding author: ayut@g.sut.ac.th},
K. Khosonthongkee$^{1,2}$, Y. Yan$^{1,2}$
}
\address{
$^{1}$\mbox{School of Physics, Institute of Science, Suranaree University of Technology,
Nakhon Ratchasima 30000, Thailand}\\
$^{2}$\mbox{Thailand Center of Excellence in Physics (ThEP), Commission on Higher Education, Bangkok 10400, Thailand} \\
}
\date{\today}

\begin{abstract}
We study the reaction $e^+e^-\to\bar DD$ near threshold in the
$^3P_0$ non-relativistic quark model, including as intermediate
states the $J/\psi$, $\psi(2S)$, $\psi(3770)$ and
$\psi(4040)$ mesons. The work reveals that
experimental data strongly favor one of the two $\psi(2S)-\psi(3770)$ mixing angles
derived by fitting to the $e^-e^+$ partial decay widths of the $\psi(2S)$ and $\psi(3770)$ mesons.
The meson $X(3940)$ as well as the resonance around 3.9 GeV
observed by Belle and BaBar Collaborations in the reaction $e^+e^-\rightarrow \bar DD$
is unlikely to be a $c\bar c$ $I^G(J^{PC})=0^-(1^{--})$ state.
\end{abstract}


\maketitle

\section{Introduction}
The recent studies of the exclusive initial state radiation production $e^+e^-\to\bar DD$ near threshold from Belle \cite{Belle1} and BaBar \cite{BaBar1, BaBar2} collaborations have consistently reported an enhancement around 3.9 GeV in the $\bar DD$ mass spectrum. The $D\bar D$ production in $e^+e^-$ annihilations near threshold is investigated in an effective Lagrangian approach \cite{Zhang}, where the $X(3900)$ is included as a $J^{PC}=1^{--}$ meson. It is concluded in \cite{Zhang} that the inclusion of the $X(3900)$ is essential to reproduce the experimental data \cite{Belle1, BaBar1, BaBar2}. As a vector charmonium with $J^{PC}=1^{--}$ at
3.9 GeV will cause great concern about the non-relativistic $\bar cc$ phenomenology, the alternative explanation of the bump structure
by the $\bar D^*D+c.c.$ open charm effects via intermediate meson loops  is investigated \cite{Zhang}.

By employing a partial reconstruction technique to increase the detection efficiency and suppress background, Belle first observed a peak around 3.94 GeV
in spectrum of mass recoiling against the $J/\psi$ in the inclusive process $e^+e^-\to J/\psi ~X$ with $X$ decaying to $\bar D D^*$ \cite{Belle2}. Later, the processes $e^+e^-\to J/\psi~\bar D^{(*)}D^{(*)}$ was studied and the observation of a charmonium-like state with mass about 3.94 GeV was confirmed \cite{Belle3}. The reaction $e^+e^-\to J/\psi\,X(3940)$ is studied in the framework of light cone formalism \cite{Braguta2006}, supposing that the $X(3940)$ is a $3^1S_0$ state or one of the $2^3P$ states. It is suggested
in \cite{Braguta2006} that the $X(3940)$ is a $3^1S_0$ charmonium. The most likely interpretation of the
$X(3940)$ is that it is the $3^1S_0\,(\bar cc)\,\eta_c(3S)$ state (see Ref. \cite{Eichten} for a recent review).

In this work, we study in the $^3P_0$ non-relativistic quark model the lineshape of the cross section reaction $e^+e^-\to\bar DD$ near threshold, including $X(3900)$ or $X(3940)$ along with the $J/\psi$, $\psi(2S)$, $\psi(3770)$ and $\psi(4040)$ as intermediate mesons.
We will show that the meson $X(3940)$ as well as the resonance around 3.9 GeV
observed by Belle and BaBar Collaborations is unlikely to be a charmonium state with $I^G(J^{PC})=0^-(1^{--})$.

\section{$e^+e^-\to\bar DD$ in $^3P_0$ quark model}
The reaction $e^+e^-\to\bar DD$ may stem from two
possible processes, namely, the one-step process where the
$e^+e^-$ pair annihilates into a virtual time-like photon, then the
virtual photon decays into a $\bar cc$ pair, and finally the
$\bar cc$ pair is dressed directly by an additional
quark-antiquark pair pumped out of the vacuum to form the
$\bar DD$ final state, and the two-step process where the
created $\bar cc$ pair first form a vector meson and then
the vector meson
decays into $\bar DD$. Theoretical works
in the $^3P_0$ quark model reveal that
the reactions $e^+e^-\to\pi\pi, \pi\omega, \bar NN$
are dominated by the two-step process at low
energies \cite{Yan2005,Kittimanapun,Yan2010}. We expect the reaction
$e^+e^-\to\bar DD$ near threshold is mainly a two-step
process, in line with our previous works and the vector meson dominance model
which is successfully applied to study the reaction $e^+e^-\to\bar DD$
in an effective Lagrangian approach \cite{Zhang}.

The transition amplitude
of the reaction $e^+e^-\to\bar DD$ in the two
step process takes the form
\begin{equation}\label{eq::1}
T=\sum_{\psi_i}\langle\bar DD| V_{\bar
qq}|\psi_i\rangle\langle\psi_i|G|\psi_i\rangle\langle\psi_i|\bar
qq\rangle\langle\bar qq|T|e^+e^-\rangle
\end{equation}
where $\psi_i$ stand for all $\bar cc$ $I^G(J^{PC})=0^-(1^{--})$ mesons
such as the $J/\psi$, $\psi(2S)$, $\psi(3770)$ and
$\psi(4040)$, and $\langle\psi_i|\bar qq\rangle$ are simply the wave
functions of the intermediate mesons. $\langle\psi_i|G|\psi_i\rangle$,
the Green function describes the propagation of the intermediate
mesons, and $\langle\bar DD|V_{\bar qq}|\psi_i\rangle$ is the
transition amplitude of the intermediate meson $\psi_i$ decaying to
the $\bar DD$ state in the $^3P_0$ nonrelativistic quark model.

The transition amplitude of the process $e^+e^-\to\psi_i$
in eq. ({\ref{eq::1}) can be easily derived in the standard
method of quantum field theory, taking the form
\begin{align}\label{eq::2}
    & T_{e^+e^-\to\psi_i}\nonumber \\\
    &=\sum_{M_sM_L}\sum_{s_qs_{\bar q}}\frac{2}{\sqrt3}~C(\frac{1}{2}\frac{1}{2}S,s_qs_{\bar
    q}M_s)C(SLJ,M_sM_LM)\nonumber\\
    &\quad\cdot\int\frac{d\vec
    p}{(2\pi)^{3/2}2E_q}~\Psi_{\psi_i}^{LM_L}(\vec p)~T_{e^+e^-\to\bar cc}
\end{align}
where $S$, $L$ and $J$ are respectively the total spin, orbital angular momentum
(either 0 or 2) and total angular momentum (actually equal to 1) of
the $\bar cc$ pair of the $\psi_i$ meson, $\Psi_{\psi_i}^{LM_L}(\vec p)$
is the spatial wave function of the $\psi_i$ meson in momentum space with
$\vec p\,$ being the relative momentum between the quark and
antiquark inside, and $T_{e^+e^-\to\bar cc}\equiv\langle
q\bar q|T|e^+e^-\rangle$ is the transition amplitude of the reaction
of $e^+e^-\rightarrow \bar qq$, taking the
form
\begin{eqnarray}\label{eq::3}
\langle e^+e^-|T|q\bar q\rangle &=& -\frac{e_qe}{s}\bar
    u_e(p_{e^-},m_{e^-})\gamma^\mu v_e(p_{e^+},m_{e^+}) \nonumber \\
    && \bar v_q(p_{\bar q},m_{\bar q})\gamma_\mu
    u_q(p_{q},m_{q})
\end{eqnarray}
where $s=(p_q+p_{\bar q})^2$, $e_q$ is the quark charge, and the
Dirac spinors are normalized according to $\bar uu=\bar vv=2m_q$.

The Green function in Eq. (\ref{eq::1})
describing the propagation of the intermediate meson takes the
form
\begin{eqnarray} \label{eq::4}
\langle\psi_i|G|\psi_i\rangle=\frac{e^{i\phi_i}}{E_{cm}-(M_{\psi_i}-i\,\Gamma_{\psi_i}/2)}
\end{eqnarray}
where $E_{cm}$ is the center-of-mass
energy of the system, $M_{\psi_i}$ and $\Gamma_{\psi_i}$ are the mass and width of
the intermediate meson $\psi_i$, and a phase factor $e^{i\phi_i}$ is added to
the amplitude of all charmonium resonances except for the $J/\psi$ \cite{BaBar1, BaBar2, Zhang}.

The transition amplitudes
for the processes $\psi_i\to\bar DD$ are derived in the $^3P_0$ quark model.
It was shown that the $^3P_0$ approach is successful in the description of hadronic
couplings.
The $^3P_0$ decay model defines the
quantum states of a quark-antiquark pair destroyed into or
created from vacuum to be $J=0, L=1, S=1$ and $T=0$.
The effective vertex in the $^3P_0$ model takes the form as in Refs. \cite{Yan2005,Kittimanapun}
\begin{align} \label{eq::5}
    V_{ij}&=\lambda\vec\sigma_{ij}\cdot(\vec p_i-\vec
    p_j)\hat{F}_{ij}\hat{C}_{ij}\delta(\vec p_i+\vec
    p_j)\nonumber\\
    &=\lambda\sum_\mu\sqrt\frac{4\pi}{3}(-1)^\mu\sigma_{ij}^\mu Y_{1\mu}
    (\vec p_i-\vec p_j)\hat F_{ij}\hat C_{ij}\delta(\vec p_i+\vec
    p_j)
\end{align}
where $\sigma_{ij}^\mu, \hat F_{ij}, \hat C_{ij},$ and $\lambda$
are respectively the spin, flavor and color operators,
and the effective coupling constant. The operations of flavor, color, and
spin operators onto a $q\bar q$ pair are
\begin{align}\label{eq::6}
\langle 0,0|\hat{F}_{ij}|\left[\bar t_i\otimes t_j\right]_{T,T_z}\rangle &= \sqrt
    2\delta_{T,0}\delta_{T_z,0}, \nonumber \\
\langle 0,0|\hat{C}_{ij}|q_\alpha^i\bar q_\beta^j\rangle &=
    \delta_{\alpha\beta}, \nonumber \\
\langle
0,0|\sigma_{ij}^\mu|\left[\bar\chi_i\otimes\chi_j\right]_{JM}\rangle&=
    (-1)^M\sqrt 2\delta_{J,1}\delta_{M,-\mu}
\end{align}
where $\chi_i (\bar\chi_i)$ and $t_i (\bar t_i)$ are the spin and flavor states of quark (antiquark),
and $\alpha$ and $\beta$ are the color indices.

In the work we approximate
the wave function of all mesons with the Gaussian form,
\begin{eqnarray} \label{eq::7}
\Psi_{nlm}(\vec p)=N_{nl}e^{-a^2p^2/2}\,L^{l+1/2}_{n}(a\,p)\,Y_{lm}(\theta,\phi)
\end{eqnarray}
where $L^{l+1/2}_{n}(x)$ are the generalized Laguerre polynomial, $\vec p$ is the
relative momentum between the quark and antiquark in a meson, and $a$ is the
length parameter of the Guassian-type wave function.
As the final state mesons are spinless, there exists only the P-wave transition amplitude
for the processes $\psi_i\to\bar DD$, that is
\begin{align}\label{eq::8}
T_{\psi_i\to\bar DD} = \sum_{m=-1}^1\,F_{n,l=1}(k)\,Y_{l=1,m}(\hat k)
\end{align}
with $F_{n,l=1}(k)$ taking the general form
\begin{align}
    F_{n,l}= A_1\,k\left(1+A_2\,k^2+A_4\,k^2\right)\,e^{-\frac{b^2 B^2 k^2}{4 \left(b^2+2 B^2\right)}}
\end{align}
where $\vec k$ is the relative momentum between the two final mesons, and $b$ and $B$ are respectively the length parameters of
the intermediate $\psi'$ meson and the
final $D(\bar D)$ meson.
For the purpose of good documentation, we
list obviously the non-zero coefficients in eq. (\ref{eq::8}) for the processes $\psi_i(nS)\to\bar DD$ and $\psi_i(nD)\to\bar DD$. We have
\begin{eqnarray}
    \psi(1S): && \frac{8 \sqrt{2} b^{3/2} B^3 \left(b^2+B^2\right)}{3 \sqrt[4]{\pi } \left(b^2+2 B^2\right)^{5/2}} \nonumber \\
    \psi(2S): && -\frac{8 b^{3/2} B^3 \left(b^2-3 B^2\right) \left(3 b^2+2 B^2\right)}{3 \sqrt{3} \sqrt[4]{\pi } \left(b^2+2 B^2\right)^{7/2}} \nonumber \\
    \psi(3S): &&  \frac{4 \sqrt{\frac{5}{3}} b^{3/2} B^3 \left(b^2-2 B^2\right) \left(3 b^4-11 b^2 B^2-6 B^4\right)}{3 \sqrt[4]{\pi } \left(b^2+2 (B^2\right)^{9/2}}\nonumber \\
    \psi(1D): &&  \frac{32 \sqrt{\frac{5}{3}} b^{7/2} B^5}{3 \sqrt[4]{\pi } \left(b^2+2 B^2\right)^{7/2}}\nonumber \\
    \psi(2D): &&  -\frac{16 \sqrt{\frac{70}{3}} b^{7/2} B^5 \left(b^2-2 B^2\right)}{3 \sqrt[4]{\pi } \left(b^2+2 B^2\right)^{9/2}}
\end{eqnarray}
for the coefficient $A_1$,
\begin{eqnarray}
    \psi(2S): && \frac{2 b^2 B^4 \left(b^2+B^2\right)}{\left(b^2-3 B^2\right) \left(b^2+2 B^2\right) \left(3 b^2+2 B^2\right)}\nonumber \\
    \psi(3S): && \frac{4 b^2 B^4 \left(5 b^4-9 b^2 B^2-10 B^4\right)}{5 \left(b^2-2 B^2\right) \left(b^2+2 B^2\right) \left(3 b^4-11 b^2 B^2-6 B^4\right)}\nonumber \\
    \psi(1D): && -\frac{B^2 \left(b^2+B^2\right)}{5 \left(b^2+2 B^2\right)}\nonumber \\
    \psi(2D): && -\frac{B^2 \left(b^4-3 b^2 B^2-2 B^4\right)}{5 \left(b^2-2 B^2\right) \left(b^2+2 B^2\right)}
\end{eqnarray}
for the coefficient $A_2$, and
\begin{eqnarray}
      \psi(3S): &&  \frac{4 b^4 B^8 \left(b^2+B^2\right)}{5 \left(b^2-2 B^2\right) \left(b^2+2 B^2\right)^2 \left(3 b^4-11 b^2 B^2-6 B^4\right)}\nonumber \\
      \psi(2D): &&  -\frac{2 b^2 B^6 \left(b^2+B^2\right)}{35 \left(b^2-2 B^2\right) \left(b^2+2 B^2\right)^2}
\end{eqnarray}
for the coefficient $A_4$.

In our first
calculation we have four intermediate mesons $J/\psi$, $\psi(2S)$, $\psi(3770)$ and
$\psi(4040)$ included with their masses and widths
taken from the particle data group \cite{PDG}. While the
$J/\psi$ is kept always as a $1S$ meson, the study of the reactions $\psi(2S)\to e^-e^+$ and
$\psi(3770)\to e^-e^+$ reveals that the $\psi(2S)$ possess
a small D-wave component \cite{Kuang,Rosner}. Let
\begin{eqnarray}\label{mixture1}
\psi(2S)=\cos\theta_1\,|2S\rangle-\sin\theta_1\,|1D\rangle \nonumber\\
\psi(3770)=\sin\theta_1\,|2S\rangle+\cos\theta_1\,|1D\rangle
\end{eqnarray}
where $\theta_1$ is the mixing angle between the $2S$ and $1D$ states. In analogy,
the $\psi(4040)$ is also allowed to be the mixture of the $3S$ and $2D$ waves, that is
\begin{eqnarray}\label{mixture2}
\psi(4040)=\sin\theta_2\,|3S\rangle+\cos\theta_2\,|2D\rangle
\end{eqnarray}

We fit our theoretical result to the experimental data from Belle and BaBar, letting
all the relative phase
factors $\phi_i$ as well as the mixing angles $\theta_1$ and $\theta_2$ and the effective coupling constant
$\lambda$ be free parameters and letting the length parameters $B$ and $b$ run in a large region from 1.0 to 5.0 GeV$^{-1}$.
We found that it is impossible to reproduce the lineshape of
the $\psi(3770)$ meson as well as the bump structure
observed around 3.9 GeV in the $e^+e^-\to\bar DD$ cross section.

It is necessary to include as the intermediate state a resonance at 3.9 GeV.
The candidate could be
the $X(3940)$ as it is not ruled out that the X(3940) may have the
quantum number $I^G(J^{PC})=0^-(1^{--})$.
The study of the reaction $e^+e^-\to\bar DD$ in an effective Lagrangian approach \cite{Zhang}
suggests a $0^-(1^{--})$ resonance with a mass about 3.9 GeV and width about 90 MeV (denoted as $X(3900)$).
We include either the $X(3900)$ or the $X(3940)$ by pairing it with the $\psi(4040)$,
\begin{eqnarray}\label{mixture3}
&& X(3900) (X(3940))=\cos\theta_2\,|3S\rangle-\sin\theta_2\,|2D\rangle \nonumber\\
&& \psi(4040)=\sin\theta_2\,|3S\rangle+\cos\theta_2\,|2D\rangle
\end{eqnarray}

We study the reaction $e^+e^-\to\bar DD$ in the
$^3P_0$ quark model with as many model parameters
as possible predetermined. The size parameter $B$ is determined with the process
$D^+\to\mu^+\nu_\mu$. The partial decay width of the reaction $D^+\to\mu^+\nu_\mu$
is evaluated with
\begin{eqnarray}
\Gamma = \frac{p_f}{32\,M_D\,\pi^2}\int |T_{D^+\to\mu^+\nu_\mu}|^2 d\Omega
\end{eqnarray}
with
\begin{eqnarray}
T_{D^+\to\mu^+\nu_\mu}=\int \frac{d\vec p}{(2\pi)^{3/2}}\psi(\vec p)\frac{\sqrt{2M_D}}{\sqrt{2E_1}\sqrt{2E_2}} T_{c\bar d\to\mu^+\nu_\mu}
\end{eqnarray}
where $T_{c\bar d\to\mu^+\nu_\mu}$ is the transition amplitude of the process $u\bar d\to\mu^+\nu_\mu$ and $\psi(\vec p)$
is the $D$ meson wave function in momentum space. Used as inputs
the weak coupling constant $G=1.166\times 10^{-5}$ GeV$^{-2}$, the CKM element $|V_{cd}|=0.230$, the $D^+$ meson mass $M_D=1.870$ GeV, the
$c$ quark mass $m_c=1.27$ GeV, the $d$ quark mass as the constituent mass $m_d=0.35$ GeV, and the experimental
value of $\Gamma_{D^+\to\mu^+\nu_\mu}=2.42\times 10^{-7}$ eV, we derive the size parameter $B$ of
the $D$ meson to be 2.28 GeV$^{-1}$. Note that it is impossible to estimate an error range for the the size parameter $B$ as
the CKM element $|V_{cd}|$ alone would lead to a sizable error bar for the $D$ meson decay width.

The size parameter $b$ of the $\psi(2S)$ and $\psi(3770)$ mesons and the mixing angle $\theta_1$ in eq. (\ref{mixture1})
are determined by the reactions $\psi(2S)\to e^-e^+$ and $\psi(3770)\to e^-e^+$ in the present model. The decay width of these
two reactions can be worked out the same way as for the process $D^+\to\mu^+\nu_\mu$. Fitting the experimental values
of $\Gamma_{\psi(2S)\to e^-e^+}=2.35\pm 0.04$ keV and $\Gamma_{\psi(3770)\to e^-e^+}=0.262\pm 0.018$ keV leads to
$b=1.95\pm 0.01$ GeV$^{-1}$ and the mixing angle $\theta_1$ being $10.69\pm 0.63^o$ or $-27.6\pm 0.69^o$ in the present model.

With $\theta_1$ being $10.69\pm 0.63^o$
or $-27.6\pm 0.69^o$, the fit of the theoretical result of the partial decay width of
the process $\psi(3770)\to \bar DD$ in the $^3P_0$ model to the experimental data, $\Gamma_{\psi(3770)\to D^+D^-}=11.15\pm 1.09$ MeV
and $\Gamma_{\psi(3770)\to D^0\bar D^0}=14.14\pm 1.36$ MeV \cite{PDG}
leads to the effective coupling
strength $\lambda=0.68\pm 0.04$ or $\lambda=4.15\pm 0.20$.

We fit the lineshape of the $\psi(3770)$ meson in the $e^+e^-\to\bar DD$ cross section with two sets of model parameters,
that is, $\{B=2.28\; {\rm GeV}^{-1},\,b=1.95\; {\rm GeV}^{-1},\,\theta_1=10.69^o,\,\lambda=0.68\}$ and
$\{B=2.28\; {\rm GeV}^{-1},\,b=1.95\; {\rm GeV}^{-1},\,\theta_1=-27.6^o,\,\lambda=4.15\}$. Here only the $J/\psi$, $\psi(2S)$ and $\psi(3770)$ are
included as the intermediate mesons. It is found that the experimental data strongly
favor the first set of parameters as the second set of parameters leads to a $\psi(3770)$
peak over 10 times higher than the data. It is also noted that the error of the size parameter $b$, $0.01\; {\rm GeV}^{-1}$, has very little effect on the cross section.

With $B=2.28$ GeV$^{-1}$ for the final $D(\bar D)$ meson, $b=1.95$ GeV$^{-1}$ for all intermediate $\psi_i$,
$10.69^o$ for the $\psi(2S)-\psi(3770)$ mixing angle and $\lambda=0.68\pm 0.04$ for the
effective coupling strength of the $^3P_0$ vertex, we fit
the Belle and BaBar data of the processes $e^+e^-\to\bar D^0D^0$ and $D^+D^-$ at energies from the $\bar DD$ threshold to 4.2 GeV,
where the errors of the experimental data
have been included in the fitting process. The $J/\psi$, $\psi(2S)$, $\psi(3770)$,
$\psi(4040)$, and $X(3900)$ or $X(3940)$ are included as the
intermediate states, and the decay width of the $X(3900)$ or $X(3940)$ and
the mixing angle $\theta_2$ as well as
all the relative phase factors $e^{-i\phi_i}$ are free parameters in the calculations.
The fitted parameters and $\chi^2/{\rm DOF}$ with regard to the central values of the parameters
are listed in Table \ref{tab1}, where the second column (Fit I) and the third (Fit II) are
from the calculations with the $X(3900)$ and $X(3940)$ included as the intermediate meson, respectively.
The theoretical results with the central values of the parameters in Table \ref{tab1} are compared with experimental data in Fig. \ref{fig1} for the
cross section of the reaction $e^+e^-\to\bar D^0D^0$.
The theoretical results for the $e^+e^-\to D^+D^-$ cross section are similar to the ones for
the reaction $e^+e^-\to\bar D^0D^0$.
The curves in the first and second panels of Fig. \ref{fig1} are the results with the $X(3900)$ and $X(3940)$ included as the intermediate state, respectively.
The decay widths of the $X(3900)$ and $X(3940)$ are fitted to be $210 \pm 19$ MeV and $268\pm 17$ MeV, respectively. A decay width of 200 MeV is much larger than
the one predicted in Ref. \cite{Zhang} for the $X(3900)$ and some three times the experimental upper limit of the $X(3940)$ decay width \cite{PDG}.
\begin{table}[h]
\begin{center}
\caption{Model parameters: Fit I (Fit II) from the calculation with $X(3900)$ ($X(3940)$) included as intermediate meson.
The $^3P_0$ coupling strength $\lambda=0.68\pm 0.04$ is input for all intermediate mesons.}\label{tab1}
\vspace*{.3cm}
\begin{tabular}{lccc}
\hline
\hline
\\
Parameters & Fit I  &&  Fit II \\
\\
\hline
\\
$\lambda_{D\bar D X}$ & $0.68 \pm 0.04$  && $0.68\pm 0.04$  \\
\\
$\Gamma_{X}$ (MeV) & $210 \pm 19$   && $268\pm 17$  \\
\\
$\theta_2$  & $28.6\pm 1.9$ && $13.4\pm 1.8$ \\
\\
$\phi_{\psi(2S)}$ & $164\pm 10$ && $175\pm 6$ \\
\\
$\phi_{\psi(3770)}$ & $55\pm 12$  && $65\pm 9$  \\
\\
$\phi_{\psi(4040)}$ & $250\pm 11$  && $257\pm 7$ \\
\\
$\phi_{X}$ & $170\pm 13$ && $170\pm 15$ \\
\\
$\chi^2/{\rm DOF}$ &0.08 &&  0.09 \\
\\
  \hline
  \hline
\end{tabular}
\end{center}
\end{table}

\begin{table}[h]
\begin{center}
\caption{Model parameters: Fit III (Fit IV) from the calculation with $X(3900)$ ($X(3940)$) included as intermediate meson.
$\Gamma_{X(3900)}=90$ MeV and $\Gamma_{X(3940)}=70$ MeV are taken from Ref. \cite{Zhang} and Ref. \cite{PDG}, respectively.}\label{tab2}
\vspace*{.3cm}
\begin{tabular}{lccc}
\hline
\hline
\\
Parameters & Fit III && Fit IV \\
\\
\hline
\\
$\Gamma_{X}$ (MeV)  & $90\pm 12$ && $70\pm 11$ \\
\\
$\lambda_{D\bar D X}$ & $0.24\pm 0.03$ && $0.15\pm 0.02$ \\
\\
$\theta_2$  & $26.9\pm 0.8$ && $19.4\pm 1.2$ \\
\\
$\phi_{\psi(2S)}$   & $164\pm 2$ && $195\pm 2$ \\
\\
$\phi_{\psi(3770)}$ & $42\pm 6$ && $71\pm 4$\\
\\
$\phi_{\psi(4040)}$  & $207\pm 14$ && $209\pm 15$ \\
\\
$\phi_{X}$ & $143\pm 12$ && $132\pm 12$ \\
\\
$\chi^2/{\rm DOF}$ & 0.10 && 0.11 \\
\\
  \hline
  \hline
\end{tabular}
\end{center}
\end{table}

\begin{figure}
\begin{center}
  \centering
  \includegraphics[width=0.4\textwidth]{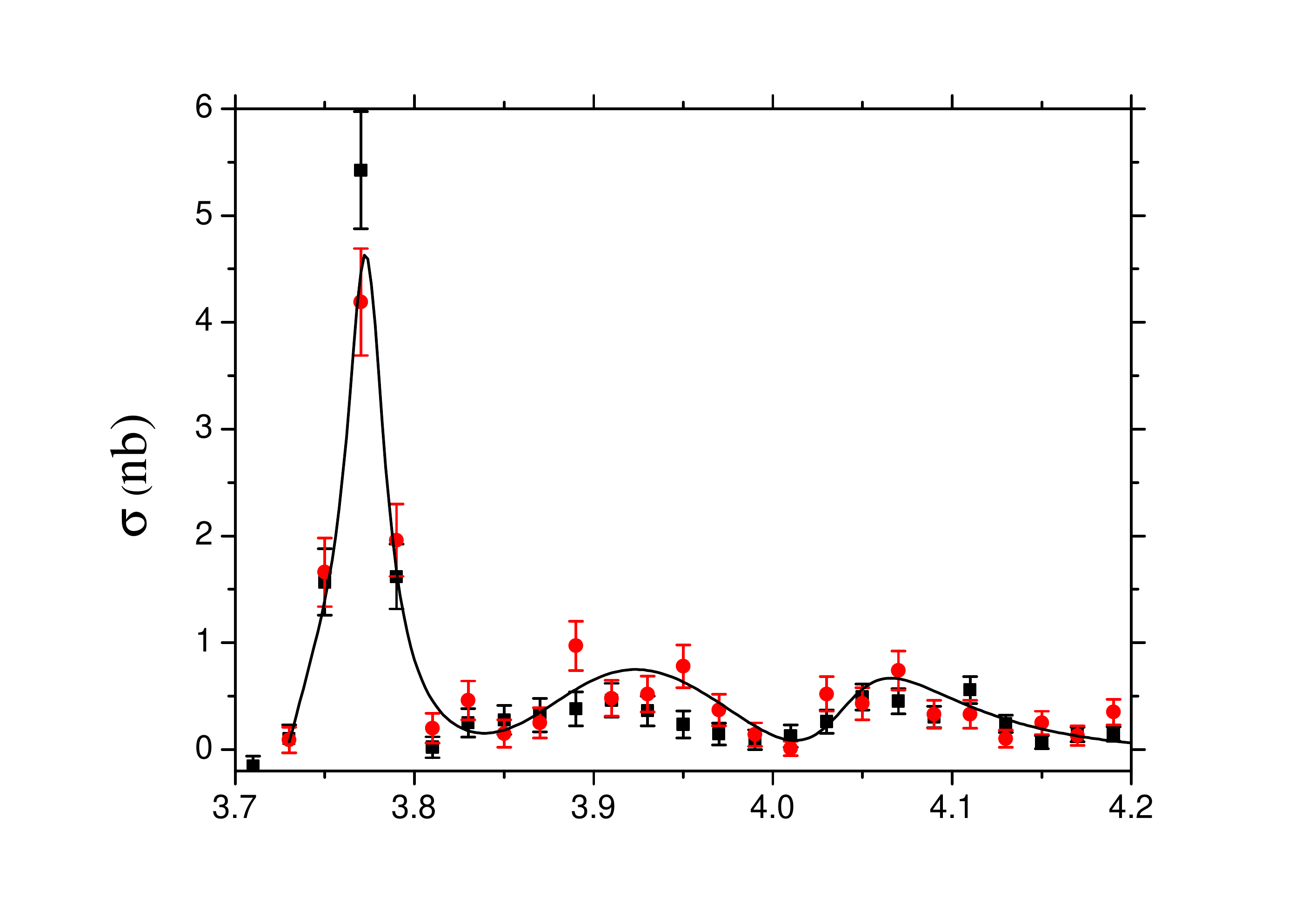} \\
  \vspace{-\baselineskip}
  \includegraphics[width=0.4\textwidth]{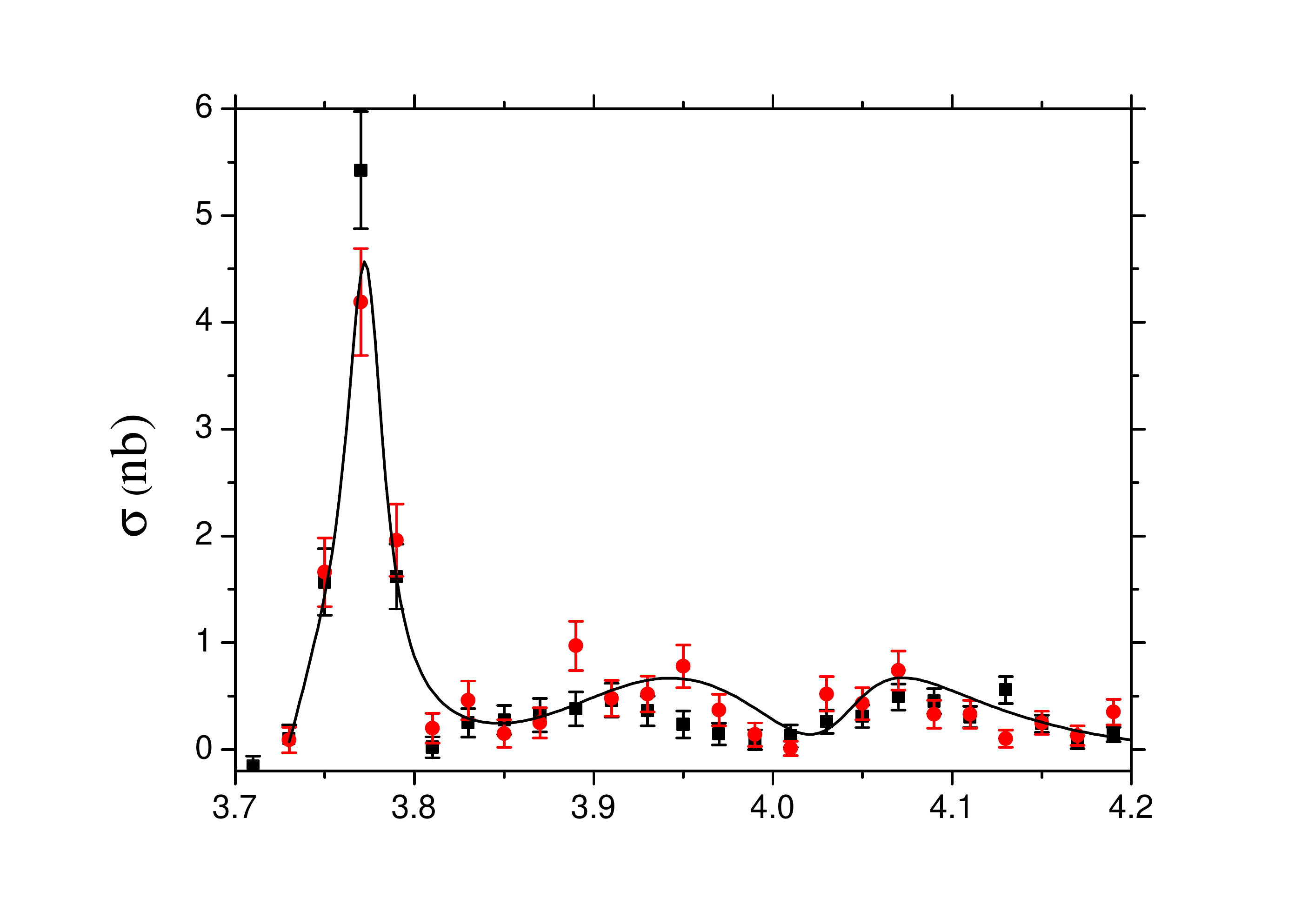} \\
  \vspace{-\baselineskip}
  \includegraphics[width=0.4\textwidth]{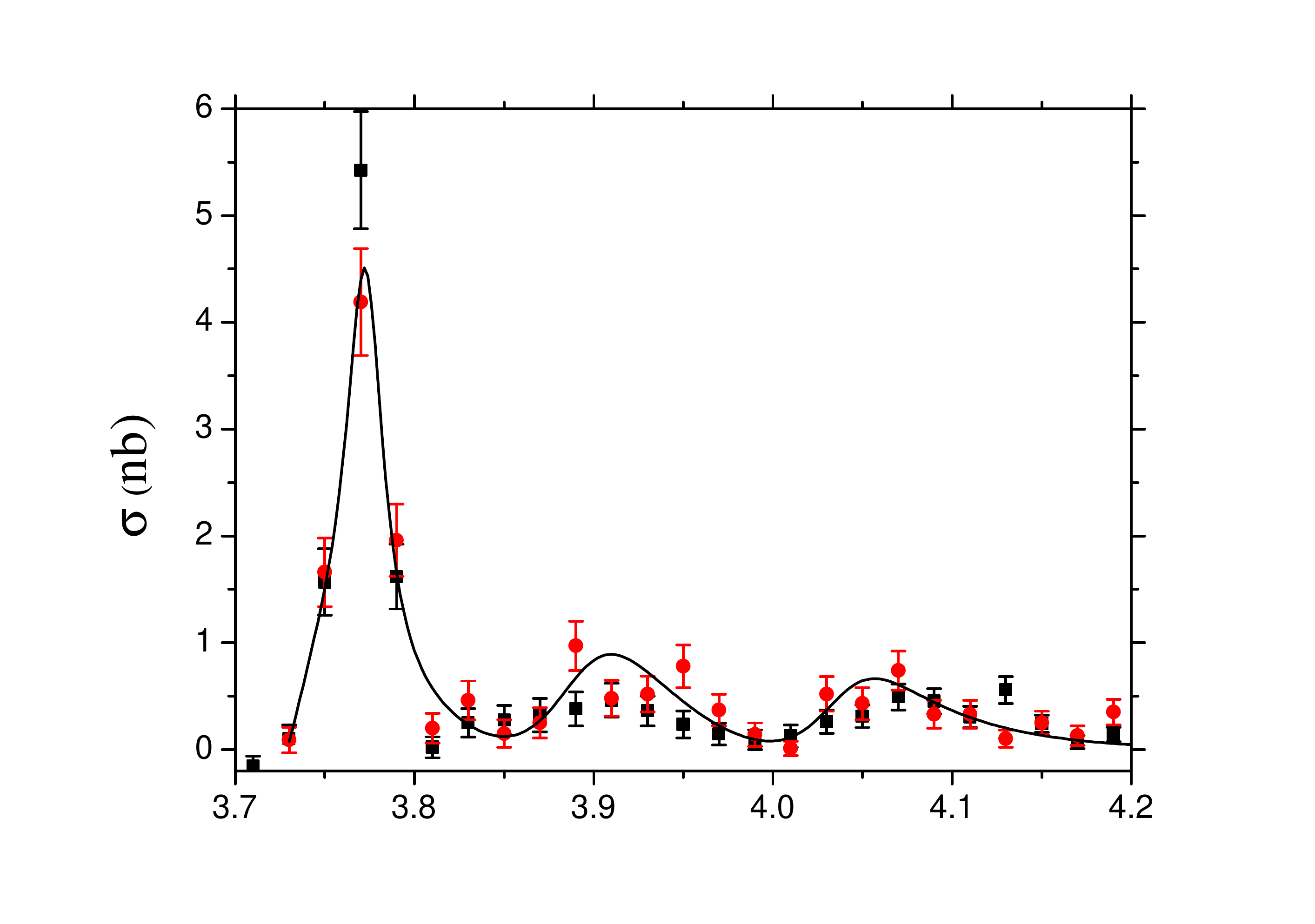} \\
  \vspace{-\baselineskip}
  \includegraphics[width=0.4\textwidth]{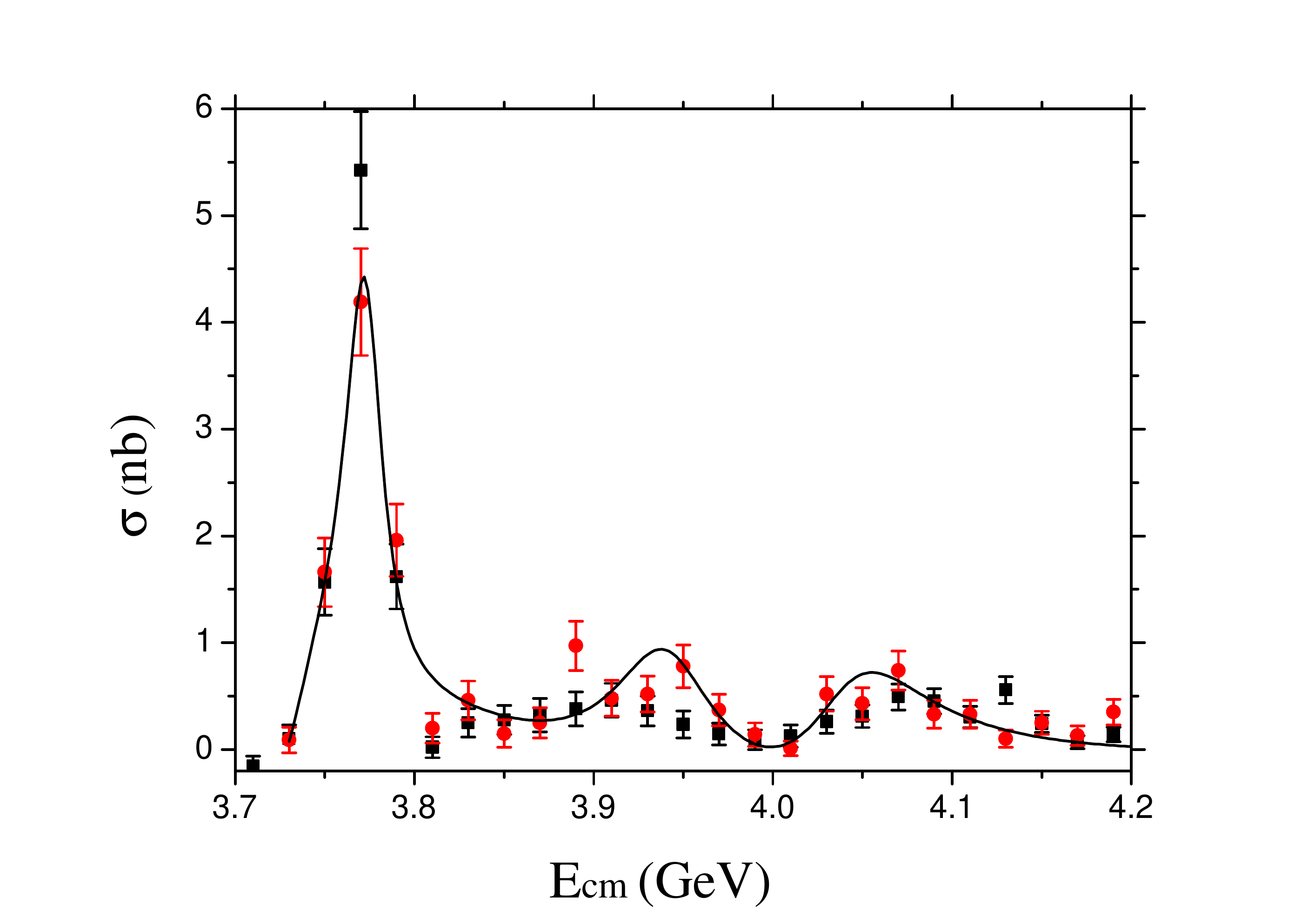}
  %\vspace{-\baselineskip}
  \caption{\label{fig1} %
  Theoretical results for the cross section of the reaction $e^+e^-\to\bar D^0D^0$. First panel: $X(3900)$
  included and its decay width a free parameter; Second panel: $X(3940)$ included and its decay width a free parameter; Third panel:  $X(3900)$
  included and the coupling strength $\lambda_{D\bar D X(3900)}$ a free parameter; Fourth panel:  $X(3940)$ included
  the coupling strength $\lambda_{D\bar D X(3940)}$ a free parameter.
    The experimental data are taken from the Belle \cite{Belle1} and the BaBar \cite{BaBar1}.}
\end{center}
\end{figure}
In the first calculation mentioned above, we have input the same $^3P_0$ vertex strength $\lambda=0.68\pm 0.04$ for all the intermediate mesons and fitted the
decay widths of the $X(3900)$ and $X(3940)$ mesons.
Instead of doing so in the second calculations, we let the $^3P_0$ vertex
strength $\lambda$ free for both the $X(3900)$ and $X(3940)$ mesons and take
the decay width
from Ref. \cite{Zhang} for the $X(3900)$ and the one from Ref. \cite{PDG} for the $X(3940)$ as input parameters.
The fitted model parameters are listed
in Table \ref{tab2}, where the second column (Fit III) and the third (Fit IV) are
from the calculations with the $X(3900)$ and $X(3940)$ included as the intermediate meson, respectively.
To see the errors of the fitted parameters, we have considered the errors of the experimental data in the fitting process.
The theoretical results with the central values of the parameters in Table \ref{tab2} are plotted in the third and fourth panels
in Fig. \ref{fig1} for the
cross section of the reaction $e^+e^-\to\bar D^0D^0$ compared with the Belle and BaBar data.
It turns out that the effective
coupling strengths of the $^3P_0$ vertex for the reactions $X(3900)\to\bar DD$ and $X(3940)\to\bar DD$ are
much smaller than the one, $\lambda=0.68$, for the decay processes $\psi(2S)\to\bar DD$, $\psi(3770)\to\bar DD$ and $\psi(4040)\to\bar DD$.

\section{Discussion and Conclusions}
The near threshold $e^+e^-\to\bar DD$ reaction is
investigated in the $^3P_0$ quark model with a number of
model parameters predetermined by other processes.
The model study reveals that it is necessary to include as the intermediate states the resonance
$X(3900)$ or $X(3940)$ as well as $J/\psi$, $\psi(2S)$, $\psi(3770)$ and
$\psi(4040)$ to reproduce the experimental data for the
$e^+e^-\to\bar DD$ cross section.

It is found that experimental data rule out one of the two $\psi(2S)-\psi(3770)$ mixing angles
derived by fitting to the $e^-e^+$ partial decay widths of the $\psi(2S)$ and $\psi(3770)$ mesons.

We have assumed that the $X(3900)$ or $X(3940)$ is a $c\bar c$ $I^G(J^{PC})=0^-(1^{--})$ state and hence applied the same coupling
strength of the $^3P_0$ vertex for the $X(3900)\to\bar DD$ and $X(3940)\to\bar DD$ decays as
for the processes $\psi(2S)\to\bar DD$, $\psi(3770)\to\bar DD$ and $\psi(4040)\to\bar DD$. By fitting to the experimental data,
however, the assumption leads to a decay width for
either the $X(3900)$ or $X(3940)$, which is much larger than the experimental data \cite{PDG} or the prediction of other work \cite{Zhang}.

Instead of using as inputs the same coupling strength for all the intermediate mesons, we have input the experimental decay width for the $X(3940)$
and the width from Ref. \cite{Zhang} for the $X(3900)$. It turns out that the experimental data of the $e^+e^-\to\bar DD$ cross section dictate a
much smaller coupling strength of the $^3P_0$ vertex for either $X(3900)$ or $X(3940)$ than the one for
the $c\bar c$ $I^G(J^{PC})=0^-(1^{--})$ states $\psi(2S)$, $\psi(3770)$ and $\psi(4040)$.

The study reveals that, without including the $X(3900)$ or $X(3940)$ as the intermediate state, it is impossible to reproduce the lineshape of
the $\psi(3770)$ meson as well as the bump structure
observed around 3.9 GeV in the $e^+e^-\to\bar DD$ cross section in the present model. We have assumed the $X(3900)$ or $X(3940)$ to be a normal 
$c\bar c$ $I^G(J^{PC})=0^-(1^{--})$ meson to fit the experimental data, but derived a much weaker coupling strength  
of the $^3P_0$ vertex for the reactions $X(3900)\to\bar DD$ and $X(3940)\to\bar DD$ than for the processes 
$\psi(2S)\to\bar DD$, $\psi(3770)\to\bar DD$ and $\psi(4040)\to\bar DD$. Therefore, one may concludes that the $X(3940)$ and $X(3900)$
are unlikely to be normal $c\bar c$ $I^G(J^{PC})=0^-(1^{--})$ states.

\section*{Acknowledgements}
This work is supported by TRF-CHE-SUT under
contract No. MRG5480186. AL would like to thank TTSF for extra support.

 \end{document}